\documentclass[aps,prl,twocolumn,superscriptaddress,showpacs]{revtex4}

\usepackage{graphicx}

\begin{document}
 
\preprint{XXXX-XX}

\title{The proton and deuteron $F_2$ structure function at low $Q^2$ \\ }
 
\author{V.~Tvaskis}
\affiliation{VU University, 1081 HV Amsterdam, The Netherlands}
\affiliation{National Institute for Subatomic Physics (Nikhef), 
1009 DB Amsterdam, The Netherlands}
\affiliation{Hampton University, Hampton, Virginia 23668}

\author{J.~Arrington}
\affiliation{Argonne National Laboratory, Argonne, Illinois 60439}

\author{R.~Asaturyan}
\affiliation{Yerevan Physics Institute, 375036, Yerevan, Armenia}

\author{O.~K.~Baker}
\affiliation{Hampton University, Hampton, Virginia 23668}

\author{H.~P.~Blok}
\affiliation{VU University, 1081 HV Amsterdam, The Netherlands}
\affiliation{National Institute for Subatomic Physics (Nikhef), 
1009 DB Amsterdam, The Netherlands}

\author{P.~Bosted}
\affiliation{Thomas Jefferson National Accelerator Facility, Newport News, Virginia 23606}

\author{M.~Boswell}
\affiliation{Randolph-Macon Woman's College, Lynchburg, Virginia 24503}

\author{A.~Bruell}
\affiliation{Massachusetts Institute of Technology, Cambridge, Massachusetts 02139}

\author{M.~E.~Christy}
\affiliation{Hampton University, Hampton, Virginia 23668}

\author{A.~Cochran}
\affiliation{Hampton University, Hampton, Virginia 23668}



\author{R.~Ent}
\affiliation{Thomas Jefferson National Accelerator Facility, Newport News, Virginia 23606}


\author{B.~W.~Filippone}
\affiliation{California Institute of Technology, Pasadena, California 91125}


\author{A.~Gasparian}
\affiliation{Hampton University, Hampton, Virginia 23668}



\author{C.~E.~Keppel}
\affiliation{Hampton University, Hampton, Virginia 23668}
\affiliation{Thomas Jefferson National Accelerator Facility, Newport News, Virginia 23606}

\author{E.~Kinney}
\affiliation{University of Colorado, Boulder, Colorado 80309}

\author{L.~Lapik\'as}
\affiliation{National Institute for Subatomic Physics (Nikhef), 
1009 DB Amsterdam, The Netherlands}


\author{W.~Lorenzon}
\affiliation{University of Michigan, Ann Arbor, Michigan 48109}


\author{D.~J.~Mack}
\affiliation{Thomas Jefferson National Accelerator Facility, Newport News, Virginia 23606}

\author{J.~Mammei}
\affiliation{Juniata College, Huntingdon, Pennsylvania 16652}

\author{J.~W.~Martin}
\affiliation{Massachusetts Institute of Technology, Cambridge, Massachusetts 02139}





\author{H.~Mkrtchyan}
\affiliation{Yerevan Physics Institute, 375036, Yerevan, Armenia}



\author{I.~Niculescu}
\affiliation{The George Washington University, Washington, D.C. 20052}


\author{R.~B.~Piercey}
\affiliation{Mississippi State University, Mississippi State, Mississippi 39762}

\author{D.H.~Potterveld}
\affiliation{Argonne National Laboratory, Argonne, Illinois 60439}




\author{G.~Smith}
\affiliation{Thomas Jefferson National Accelerator Facility, Newport News, Virginia 23606}

\author{K.~Spurlock}
\affiliation{Mississippi State University, Mississippi State, Mississippi 39762}

\author{G.~van der Steenhoven}
\affiliation{National Institute for Subatomic Physics (Nikhef), 
1009 DB Amsterdam, The Netherlands}

\author{S.~Stepanyan}
\affiliation{Yerevan Physics Institute, 375036, Yerevan, Armenia}

\author{V.~Tadevosian}
\affiliation{Yerevan Physics Institute, 375036, Yerevan, Armenia}



\author{S.~A.~Wood}
\affiliation{Thomas Jefferson National Accelerator Facility, Newport News, Virginia 23606}



\date{\today}

\begin{abstract}
Measurements of the proton and deuteron $F_2$ structure functions are presented.
The data, taken at Jefferson Lab Hall C, span the four-momentum transfer range
$0.06 < Q^2 < 2.8$ GeV$^2$, and Bjorken~$x$ values from 0.009 to 0.45,
thus extending the knowledge of $F_2$ to low values of $Q^2$ at low $x$.
Next-to-next-to-leading  order calculations using recent parton distribution
functions start to deviate from the data for $Q^2<2$ GeV$^2$ at the
low and high $x$-values.
Down to the lowest value of $Q^2$, the structure function is in good
agreement with a parameterization of $F_2$  based on data that
have been taken at much higher values of $Q^2$ or much lower values of $x$,
and which is constrained by data at the photon point.
The ratio of the deuteron and proton structure functions at low $x$ remains
well described by a logarithmic dependence on $Q^2$ at low $Q^2$.
\end{abstract}


\maketitle

Deep-inelastic scattering (DIS) remains a powerful tool to study the partonic
substructure of the nucleon. Decades of experiments with high-energy electron
and muon beams have provided a detailed map of the nucleon structure function
$F_2(x, Q^2)$ over many orders of magnitude in $x$ and $Q^2$ \cite{pdg}.
Here, $Q^2$ is the negative square of the four-momentum transfer of the
virtual photon exchanged in the scattering process, and $x=Q^2/2M\nu$ is the
Bjorken scaling variable, with $M$ the nucleon mass and $\nu$ the energy
of the virtual photon in the target rest frame.
In the region of large $Q^2$ and $\nu$, the results of these DIS
measurements are typically interpreted in terms of partons (quarks and gluons),
where $x$ can be interpreted as the fraction of the nucleon momentum carried by
the struck parton.

In this regime, a rigorous theoretical framework is provided by perturbative
Quantum Chromodynamics (pQCD), which gives logarithmic scaling violations
in $Q^2$ \cite{parton}. However, this description starts to fail when
non-perturbative effects such as interactions between the 
quark struck in the scattering process  and other quarks or gluons in the nucleon
become important. The sensitivity for such higher-twist effects increases with
decreasing $Q^2$, since they are proportional to powers of $1/Q^2$. Therefore,
the interpretation of the nucleon structure functions in terms of partons was
originally anticipated to become suspect when momentum and energy transfers
get below  a few GeV. Nonetheless, the perturbative descriptions
were shown to hold down to surprisingly low values of $Q^2$, of the order of 1 GeV$^2$,
provided that the energy transfer remained sufficiently large~\cite{grv}.

For small values of the energy transfer, corresponding to an invariant mass
$W = \sqrt{M^2 + 2M\nu - Q^2} < 2$ GeV of the hadronic system, the data
at low $Q^2$ prominently show excitation of nucleon resonances, and a
simple partonic interpretation fails. Moving beyond this resonance
region, the behaviour of the nucleon structure functions in
the low $Q^2$ region is thought to shed light on the transition from
perturbative to non-perturbative QCD within the partonic interpretation.
However, little is known about this behaviour, since for $W>2$ GeV there
are few data points at low $Q^2$, except for the (transverse) cross section
$\sigma_T$ at exactly $Q^2 = 0 $ from real-photon absorption experiments,
some data from SLAC~\cite{r1990}, and data at very low $x$ values ($x < 0.005$)
from the E665, ZEUS and H1 experiments~\cite{flab,zeusbpc,h1svt}.
Here, we report on measurements in the range $0.009 < x < 0.45$,
approaching the valence-quark region, for $0.06 < Q^2 < 2.8$ GeV$^2$.

The differential cross section for inclusive electron scattering can be written as
\begin{equation}
\label{sigma1}
{d^2\sigma \over d\Omega dE'} = \Gamma_v (\sigma_{T} + \varepsilon \sigma_{L} ),
\end{equation}
where the virtual-photon flux factor is given by
\begin{equation}
\label{gamma}
 \Gamma_v = {\alpha \over 2\pi^2} {E' \over E} {K \over Q^2} {1 \over 1-\varepsilon} ,
\end{equation}
with $K=(W^2-M^2)/2M$, $\varepsilon$ the virtual photon polarization, and 
$\sigma_{L}$ ($\sigma_{T}$) the longitudinal (transverse) virtual-photon
absorption cross section, which depends on $x$ and $Q^2$. 
Usually, the cross section is written in terms of the 
structure functions $F_1(x,Q^2)$ and $F_2(x,Q^2)$, where $F_1$ is proportional
to $\sigma_T$ and $F_2$ is proportional to $\sigma_L+\sigma_T$.
However, the cross section can also be written in terms of $F_2$ and
the ratio $ R \equiv \sigma_L / \sigma_T $ according to:
\begin{equation}
\label{sigma2}
 {d^2\sigma \over d\Omega dE'} = {4\pi^2\alpha \over 1-x} {1 \over Q^2}
 (1+ {4M^2x^2 \over Q^2}) \, F_2 \, {1+ \varepsilon R \over 1+R}.
\end{equation}

This equation shows that in the limit $\varepsilon \rightarrow 1$, the
structure function $F_2$ can directly be determined from the measured cross section.
Otherwise, measurements have to be performed for at least two different
values of $\varepsilon$ (beam energies) at fixed values of $x$ and $Q^2$
(Rosenbluth separation), or a value for $R$ has to be assumed.
Although results for the $F_2$ structure function have been widely
reported~\cite{r1990, rbcdms,ben2,dasu3,tao,yl1,remc,rnmc,flab,
desy_h1,desy_h2,desy_h3,zeusbpc,h1svt,zeus1,zeus2},
in many cases assumptions on the value of $R$ were made.
A more limited set of experiments actually performed
Rosenbluth-separations~\cite{r1990,rbcdms,dasu3,tao,yl1,remc,rnmc}.
In this paper we present results for $F_2$ based on data for both hydrogen
and deuterium at low values of $Q^2$, utilizing both techniques:
using Rosenbluth-separated data and using unseparated data in combination
with a parametrization of $R(x,Q^2)$.
Values of $R$ extracted from these data for those kinematics, where data
were taken at more than one value of $\varepsilon$, were reported 
previously \cite{tva1}. Here the full data set is used to determine $F_2$.

The experiment (E99-118) was
carried out in experimental Hall C at the Thomas Jefferson National
Accelerator Facility (JLab). Data were obtained for $ 0.009 < x <  0.45 $, 
$ 0.06 < Q^2 < 2.8 $ GeV$^2$ by utilizing 2.301, 3.419 and 5.648 GeV
electron beams at a current of $I = 25 \, \mu$A.  The minimum scattered electron 
energy was $E' \approx 0.4$ GeV and the range of the invariant mass 
of the hadronic system $W$ was between 1.9 and 3.2 GeV$^2$. Electrons
scattered from 4 cm long liquid hydrogen and deuterium targets
were detected in the High-Momentum  Spectrometer (HMS) in Hall C
at various angles between $10^{\circ}$ and $60^{\circ}$. 

The inclusive double differential cross section for each energy and angle bin 
within the spectrometer acceptance was determined from the measured electron
yields. The yields were corrected for detector inefficiencies, background
events, and finally radiative effects to obtain the Born cross section.
Internal bremsstrahlung, vertex corrections and loop diagrams were calculated
using the approach by Bardin~{\it et al.}~\cite{bardin}.
The code used includes the possibility that the electron emits two hard photons
($\alpha^2$ term). However, the calculation of this process is not yet fully
established.  Because the calculation including the $\alpha^2$ term overestimates
the radiative tail, perhaps because higher-order terms associated with the
emission of more than two hard photons are not negligible,
only half of the two hard-photon correction was applied, and the size of the
correction was included in the cross section systematic
uncertainty (see Ref.~\cite{tvaskis} for more details).
Additional radiative effects in the target and its exit windows were
determined using the formalism of Mo and Tsai~\cite{motsai}. 

For every bin the cross section was corrected for the variation of the
cross section over the acceptance with the angle $\theta$ to yield the
value of the cross section at the central angle (bin-centering correction).
To minimize the dependence on the model used to describe this variation
and the radiative effects, an iterative procedure was employed.
A similar procedure was used to center the cross sections at chosen
values of $x$.
For details regarding the analysis and the standard Hall~C apparatus
employed in this experiment, see Refs.~\cite{tva1,tvaskis,christy,fpi}.

The total uncertainty in the cross sections was calculated as the
quadratic sum of statistical and systematic uncertainties. 
The statistical uncertainty was in most cases well below 1\%.
The systematic uncertainty on the cross sections from instrumental sources
such as target thickness, charge integration, various efficiencies, and
acceptance amounted to 1.3\% - 1.7\%.
The uncertainty in the radiative corrections is about 1\%, except at low
values of $E^\prime$ ($E^\prime<0.8$ GeV), where the measured data are
dominated by events from elastic or quasi-elastic scattering
with the emission of one or more photons in the initial or final state.
The estimate of these uncertainties was determined by varying all relevant
input cross sections within their uncertainties, and amounted to 1.5\%
for hydrogen and 8.5\% for deuterium in the most extreme
cases considered.  The much larger uncertainty
in the deuterium cross section is due to the contribution from quasielastic
scattering, which can only be modelled approximately due to the lack of
low-$Q^2$ ($<$ 0.4 GeV$^2$) electron-deuteron scattering data over a
sufficiently wide range of energy transfers.
In addition, there is the uncertainty from the $\alpha^2$ term, which can
be as large as 50\% in a few cases (low values of $x$ and $\varepsilon$).

\begin{figure}
\includegraphics[width=3.75in,bb=0 0 550 517]{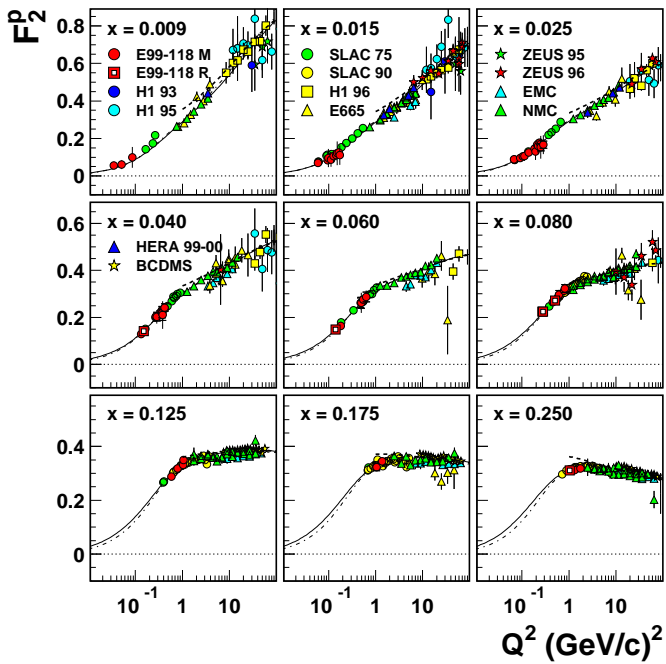}
\caption{\label{f2p}Comparison of the results for $F_2^{p}(x, Q^2)$
from the  present experiment (E99-118) to the results of other
experiments. Both the Rosenbluth-separated data (open red squares)
and the model-dependent extracted data (red circles) are shown. The thick
dashed curve represents the result of a NNLO calculation based upon the MRST
parton distributions~\protect\cite{mrst} including target-mass effects.
The solid and dot-dashed curves are the results of the phenomenological
parameterizations from Refs.~\protect\cite{f2all,gd07},
respectively.}
\end{figure}

\begin{figure}
\includegraphics[width=3.75in,bb=0 0 550 517]{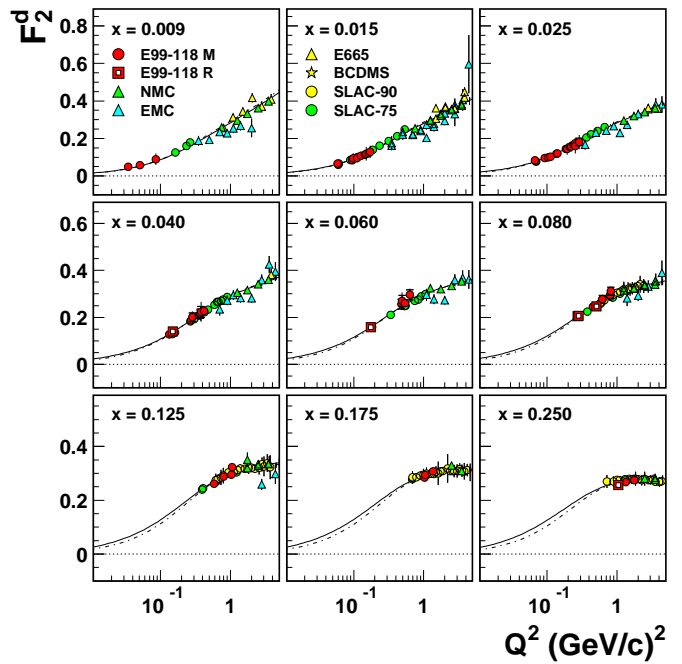}
\caption{\label{f2d}Comparison of the results for $F_2^{d}(x, Q^2)$
from the  present experiment (E99-118) to the results of other
experiments. Both the Rosenbluth-separated data (open red
squares) and the model-dependent extracted data (red circles) are
shown. The solid and dot-dashed curves result from multiplying
the phenomenological parameterizations shown in Fig.~\ref{f2p}
with ($1 + F_2^{n}/F_2^{p})/2$ using the NMC parameterization~\protect\cite{arne2} 
for the ratio $F_2^{n}/F_2^{p}$.}
\end{figure}

The results for the $F_2(x,Q^2)$ structure function for protons (deuterons)
from the Rosenbluth separated data are shown as the open squares in
Fig.~\ref{f2p} (Fig.~\ref{f2d}) as a function of $Q^2$ for fixed values of $x$.
The value of $F_2$ and its uncertainty are in essence the result of an extrapolation
of the cross sections and their (total) uncertainties, measured at the different
$\varepsilon$ values, to $\varepsilon=1$ and were calculated accordingly.
The numerical values~\cite{vtmail} are given in Table~\ref{f2_hd_ros}.

Values of $F_2$ from the unseparated data were determined by inverting
Eq.~\ref{sigma2} and using a recent parametrization of $R$.
Measurements of $R(x, Q^2)$ from the present experiment reported in~\cite{tva1}
showed a nearly constant behaviour of $R$ down to $Q^2$ of about 0.1 GeV$^2$
at low values of $x$. This was contrary to expectations that $R$ would
decrease strongly at such low $Q^2$. The unexpected behaviour
was taken into account by extending the parameterization of Ref.~\cite{r1990}
to $Q^2=0$, resulting in a new $R_{e99118}(x,Q^2)$ parameterization~\cite{tva1}.
This parameterization was used to calculate $F_2$ for all values of $x$ and $Q^2$
where cross sections from the present experiment are available
that were not used already for the Rosenbluth separation.
The uncertainty in $F_2$ is a combination of the uncertainties
in the measured cross section and in the parameterization of $R$.
For $Q^2>1 $ GeV$^2$ the latter uncertainty was taken to be 0.075 .
For $Q^2<1 $ GeV$^2$ the value of $R$ is less well known, as there are
few measurements in that region, especially at very small values of $Q^2$.
Therefore, in this region an uncertainty increasing from 0.075 at $Q^2=1 $ GeV$^2$
to 0.20 at $Q^2=0 $ GeV$^2$ was taken.
This gives a range for $R$ that covers well the existing data in this region.
The influence of $R$ on the extracted value of $F_2$ diminishes
when $\varepsilon \rightarrow 1$.
The results are shown as the red circles in Figs.~\ref{f2p},\ref{f2d}.
They agree well with the Rosenbluth-separated results, but cover
a much larger kinematic range. 
Also, there is a very good consistency between the results at the same
value of $x$ and almost the same value of $Q^2$, but different values
of $\varepsilon$, as can be seen from the numerical values given in
Table~\ref{f2_hd1_mod}.

Both figures also show the results of previous measurements at
SLAC~\cite{r1990}, by the EMC~\cite{remc}, NMC~\cite{rnmc,arne3,arne2} and
BCDMS~\cite{rbcdms,ben2} collaborations at CERN, the E665~\cite{flab}
collaboration at Fermilab, and the H1~\cite{desy_h1,desy_h2,desy_h3}
and ZEUS~\cite{zeus1,zeus2} collaborations at DESY.
In the region of $Q^2$ where these overlap with the present results there is good agreement.
At low $x$ our data clearly extend the knowledge of $F_2$ down to much lower $Q^2$.

The thick dashed curves shown in Fig.~\ref{f2p} are the result of a next-to-next-to-leading
order (NNLO) calculation based on the recent MRST parton distributions \cite{mrst},
where target-mass effects have been included according to \cite{tme}.
The calculations do not extend below $Q^2 = 1 $ GeV$^2$, as there a DIS approach is
not assumed to be applicable anymore. The calculated results closely coincide with
the data for $x \approx 0.1$. Deviations can be noticed at the lower $Q^2$ end
for the lowest ($x = 0.009$) and highest ($x = 0.250$) $x$-values.
This could be due to uncertainties in the used parton distribution functions
at low $x$ and the influence of higher-twist effects.

The solid and dot-dashed curves shown in Fig.~\ref{f2p} represent
two existing parameterizations down to the photon point of the
world's $F_2^p$ data, termed ALLM97 and GD07.
The ALLM97 parameterization \cite{f2all} represents a 23-parameter fit to
world's electron-proton scattering total cross section data based upon a
Regge-motivated approach. It includes both Reggeon and Pomeron exchange
mechanisms. Deep-inelastic scattering data covering a wide range in $x$ and
$Q^2$ are used, with additional constraints built in to connect smoothly to
the photon point.
The GD07 parameterization \cite{gd07} includes recent data and converts
from total cross sections to $F_2$ structure functions by using the 
parameterization of $R$ from Ref.~\cite{r1998} for all data.

The solid (dot-dashed) curves shown on Fig.~\ref{f2d} were constructed by
utilizing the mentioned ALLM97 and GD07 parameterizations of $F_2^p$,
multiplying them by the factor $(1 + F_2^{n}/F_2^{p})/2$ using a global
fit to $F_2^{n}/F_2^{p}$ data by the NMC collaboration~\cite{arne2}.
As can be seen from Figs.~\ref{f2p} and ~\ref{f2d}, the existing parameterizations
are in very good agreement with the measured structure functions, with deviations
often less than 3\%  and always consistent given the error bars.
The largest deviations occur for data that have large systematic uncertainties.
Given that for $x<0.1$ there were no data for $Q^2 \lesssim 1$
GeV$^2$, the agreement between our data and the
parameterizations is remarkable.  This indicates
that the constraints imposed on the parameterizations by 
low-$Q^2$ data at much smaller $x$ and the way the transition to
the real photon point at $Q^2 = 0 $ is parametrized, seem to be sufficient
to correctly predict $F_2(x,Q^2)$ in a region where it was hitherto unknown.

The ratio of the deuteron to proton structure functions, $F_2^d/F_2^p$ is
of interest because it embodies information on the neutron structure
function $F_2^n$. This ratio can be expressed in terms of the cross
section ratio $\sigma^d/\sigma^p$ as
\begin{center}
\begin{equation}
{ F_2^d \over F_2^p}
= \left({\sigma^d \over \sigma^p}\right) { (1+\varepsilon R^p)(1+R^d) 
\over (1+R^p)(1+\varepsilon R^d)} \label{eq:sf_con} \phantom{l},
\end{equation}
\end{center}
with $\sigma^d/\sigma^p$ the measured cross section ratio and 
$R^d$ and $R^p$ both functions of $x$ and $Q^2$.

\begin{figure}
\includegraphics[width=4.0in]{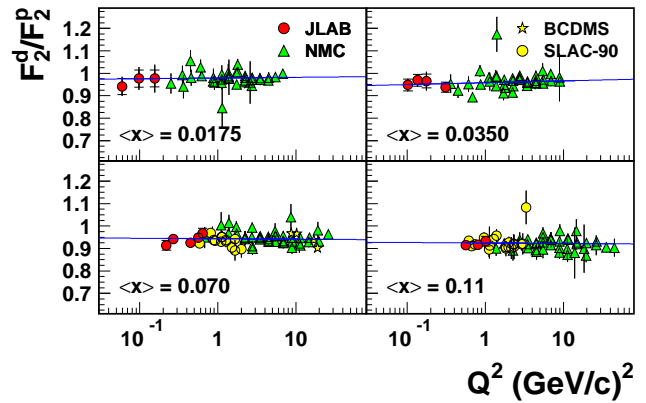}
\caption{\label{dh_q2_dep}
$Q^2$ dependence of the ratio $F^d_2/F^p_2$ for some selected $x$-bins. 
Both the present data and data from previous experiments are shown.
The lines are the results of a linear (in $\ln Q^2$) fit of all data
described in the text.}
\end{figure}

For sufficiently high energy experiments, $\varepsilon \rightarrow 1$, and
the deuteron to proton cross section and $F_2$ ratios will equate.
Similarly, the ratios equate if $R^p$ = $R^d$, which has been found
in all previous higher-$Q^2$ data, albeit with relatively large
uncertainties, see Ref.~\cite{tao,arne2}.
However, the data from the present experiment are at $\varepsilon \ne 1 $ and,
as reported earlier \cite{tva1}, there may be a small 
reduction of $R^d$ with respect to $R^p$ at low $Q^2$ ($Q^2< 1.5$ GeV$^2$).
Such a reduction was taken into account when converting our measured
$\sigma^d/\sigma^p$ ratios to $F_2^d/F_2^p$ ratios.
(The systematic error in the latter ratio includes the effect of an
uncertainty in the value of $R_d-R_p$, which conservatively was taken
equal to the value of $R_d-R_p$.)
The effect is small, decreasing the extracted ratio $F_2^d/F_2^p$ by up to a few
percent as compared to the $\sigma^d/\sigma^p$ ratio,
depending on the value of $\varepsilon$ and $Q^2$.
Fig.~\ref{dh_q2_dep} shows the present $F^d_2/F^p_2$ data for selected $x$ bins
in comparison with the world's data.
The numerical values of all present data are given in Table~\ref{ratio_dp}.

The precise data from the NMC collaboration ~\cite{arne2} have shown that the
$F^d_2/F^p_2$ ratio depends (at fixed $x$) logarithmically on the scale, $Q^2$. 
The  data from the present experiment extend the $Q^2$ range
of the data in the region $0.01< x <0.1$, thus improving the knowledge of
these logarithmic $Q^2$-dependences in that region.
Neglecting differences in higher-twist effects in the proton and deuteron,
all data for $F^d_2/F^p_2$ were fitted with the linear function 
${F^d_2 \over F^p_2}(x,Q^2) = A(x) + B(x)\ln(Q^2)$.
The data for $F_2^d/F_2^p$ from the present experiment are found to be in
excellent agreement with this parameterization, see the lines in Fig.~\ref{dh_q2_dep}.

In summary, we present $F_2$ structure function data taken on hydrogen and deuterium
spanning the four-momentum transfer range $ 0.06 < Q^2 < 2.8 $ GeV$^2$.
The data are at lower $Q^2$ values than hitherto reported in the range of
small to intermediate $x$,  $0.009 < x < 0.45 $. The data agree well with
the results of phenomenological parameterizations based upon data
accumulated at either much larger $Q^2$, or far smaller values of $x$.
Results of NNLO calculations agree with the data for $x \approx 0.1$, but deviate 
at lower and higher values of $x$, when $Q^2$ drops below 2 GeV$^2$.
The present data extend the data set for the ratio of deuteron to proton $F_2$ structure
functions to lower $Q^2$. The new data are in excellent agreement with higher-$Q^2$ data,
when the logarithmic dependence on $Q^2$ is taken into account.

\begin{acknowledgments}
This work is supported in part by research grants from the U.S. Department
of Energy, the U.S. National Science Foundation, and the Stichting voor
Fundamenteel Onderzoek der Materie (FOM) of the Netherlands.
The Southeastern Universities Research Association operates the
Thomas Jefferson National Accelerator Facility under the U.S. Department
of Energy contract DEAC05-84ER40150.
\end{acknowledgments}

\newpage

\begin{widetext}

\begin{table}
\caption{\label{f2_hd_ros}
Data for the proton and deuteron $F_2$ structure function as determined
via the Rosenbluth separation method.
Uncertainties are shown without ($\Delta_{norc}$) and with ($\Delta_{full}$)
the contribution from radiative corrections.} 
\begin{tabular}{c c c c c c c c c}
$x$  & $Q^2$  & $F_2^{p}$ & $\Delta_{norc}^{p}$  &  $\Delta_{full}^{p}$ & $F_2^{d}$ & $\Delta_{norc}^{d}$  &  $\Delta_{full}^{d}$ \\
\hline
0.040	    &  0.150   &  0.1407   &  0.0035  &  0.0046  & 0.1405  & 0.0036  &  0.0046  \\
0.060	    &  0.145   &  0.1477   &  0.0033  &  0.0053  & 0.1579  & 0.0034  &  0.0042  \\
0.080	    &  0.273   &  0.2238   &  0.0039  &  0.0053  & 0.2045  & 0.0039  &  0.0044  \\
0.080	    &  0.283   &  0.2250   &  0.0041  &  0.0054  & 0.2068  & 0.0038  &  0.0043  \\
0.080	    &  0.508   &  0.2715   &  0.0046  &  0.0074  & 0.2470  & 0.0042  &  0.0058  \\
0.175	    &  0.476   &  0.2753   &  0.0060  &  0.0062  & 0.2438  & 0.0119  &  0.0121  \\
0.250	    &  1.045   &  0.3103   &  0.0054  &  0.0057  & 0.2565  & 0.0042  &  0.0045  \\
0.350	    &  1.300   &  0.2674   &  0.0035  &  0.0036  & 0.2274  & 0.0051  &  0.0052  \\
0.350       &  1.670   &  0.2648   &  0.0046  &  0.0046  & 0.2140  & 0.0033  &  0.0038  \\
\end{tabular}
\end{table}
 
\begin{table}
\caption{\label{f2_hd1_mod}
Data for the proton and deuteron cross sections and $F_2$ structure functions as
determined via the model-dependent method.  The cross sections do not include
the virtual-photon flux factor $\Gamma_v$ (see Eqs.~\ref{sigma1},\ref{gamma}).
Uncertainties are shown without ($\Delta_{norc}$) and with ($\Delta_{full}$)
the contribution from radiative corrections.} 
\begin{tabular}{c c c c c c c c c c c c c c c c}
$x$  & $Q^2$ & $\varepsilon$ & $\sigma^p$ & $\Delta_{norc}^p$ &  $\Delta_{full}^p$ & $F_2^p$ & $\Delta_{norc}^p$ &  $\Delta_{full}^p$   & $\sigma^d$ & $\Delta_{norc}^d$ &  $\Delta_{full}^d$ & $F_2^d$ & $\Delta_{norc}^d$  &  $\Delta_{full}^d$ \\
\hline
  0.009 & 0.034 & 0.364 & 1.0493 & 0.0689 & 0.2085 & 0.0560 & 0.0072 & 0.0127 & 0.9381 & 0.0411 & 0.1287 & 0.0492 & 0.0060 & 0.0088 \\
  0.009 & 0.051 & 0.250 & 0.4638 & 0.0409 & 0.1754 & 0.0616 & 0.0094 & 0.0245 & 0.4489 & 0.0240 & 0.1121 & 0.0580 & 0.0082 & 0.0164 \\
  0.009 & 0.086 & 0.157 & 0.2382 & 0.0242 & 0.1282 & 0.0997 & 0.0166 & 0.0553 & 0.2233 & 0.0138 & 0.0694 & 0.0896 & 0.0137 & 0.0305 \\
  0.015 & 0.059 & 0.361 & 0.4304 & 0.0241 & 0.0692 & 0.0696 & 0.0082 & 0.0133 & 0.4232 & 0.0155 & 0.0456 & 0.0669 & 0.0077 & 0.0102 \\
  0.015 & 0.095 & 0.345 & 0.2711 & 0.0144 & 0.0421 & 0.0842 & 0.0093 & 0.0154 & 0.2754 & 0.0096 & 0.0301 & 0.0831 & 0.0091 & 0.0125 \\
  0.015 & 0.098 & 0.178 & 0.2034 & 0.0190 & 0.0990 & 0.0961 & 0.0150 & 0.0483 & 0.2056 & 0.0111 & 0.0549 & 0.0935 & 0.0133 & 0.0278 \\ 
  0.015 & 0.112 & 0.202 & 0.1650 & 0.0144 & 0.0712 & 0.0876 & 0.0128 & 0.0392 & 0.1893 & 0.0090 & 0.0417 & 0.0966 & 0.0130 & 0.0245 \\
  0.015 & 0.127 & 0.229 & 0.1780 & 0.0123 & 0.0514 & 0.1058 & 0.0137 & 0.0327 & 0.1878 & 0.0077 & 0.0318 & 0.1073 & 0.0134 & 0.0221 \\
  0.015 & 0.144 & 0.259 & 0.1685 & 0.0099 & 0.0372 & 0.1114 & 0.0131 & 0.0271 & 0.1773 & 0.0066 & 0.0244 & 0.1129 & 0.0131 & 0.0199 \\
  0.015 & 0.151 & 0.144 & 0.0828 & 0.0079 & 0.0421 & 0.1216 & 0.0186 & 0.0635 & 0.0846 & 0.0047 & 0.0239 & 0.1186 & 0.0166 & 0.0368 \\
  0.015 & 0.164 & 0.293 & 0.1713 & 0.0085 & 0.0270 & 0.1253 & 0.0133 & 0.0230 & 0.1742 & 0.0058 & 0.0196 & 0.1227 & 0.0131 & 0.0186 \\
  0.015 & 0.172 & 0.163 & 0.0677 & 0.0062 & 0.0305 & 0.1118 & 0.0163 & 0.0519 & 0.0816 & 0.0039 & 0.0182 & 0.1286 & 0.0169 & 0.0327 \\
  0.025 & 0.067 & 0.537 & 0.8077 & 0.0278 & 0.0482 & 0.0883 & 0.0070 & 0.0082 & 0.7744 & 0.0198 & 0.0474 & 0.0834 & 0.0066 & 0.0081 \\
  0.025 & 0.092 & 0.357 & 0.2380 & 0.0106 & 0.0280 & 0.0960 & 0.0101 & 0.0145 & 0.2422 & 0.0074 & 0.0203 & 0.0953 & 0.0101 & 0.0126 \\
  0.025 & 0.104 & 0.485 & 0.4752 & 0.0164 & 0.0327 & 0.1040 & 0.0084 & 0.0104 & 0.4636 & 0.0120 & 0.0304 & 0.0994 & 0.0082 & 0.0101 \\
  0.025 & 0.113 & 0.247 & 0.1551 & 0.0090 & 0.0360 & 0.1069 & 0.0133 & 0.0275 & 0.1534 & 0.0058 & 0.0220 & 0.1024 & 0.0125 & 0.0189 \\
  0.025 & 0.140 & 0.483 & 0.3115 & 0.0099 & 0.0173 & 0.1251 & 0.0094 & 0.0110 & 0.3075 & 0.0075 & 0.0171 & 0.1208 & 0.0093 & 0.0111 \\
  0.025 & 0.186 & 0.331 & 0.1828 & 0.0072 & 0.0207 & 0.1469 & 0.0138 & 0.0208 & 0.1849 & 0.0052 & 0.0171 & 0.1441 & 0.0138 & 0.0187 \\
  0.025 & 0.195 & 0.185 & 0.0708 & 0.0048 & 0.0234 & 0.1312 & 0.0166 & 0.0456 & 0.0807 & 0.0032 & 0.0145 & 0.1439 & 0.0173 & 0.0307 \\
  0.025 & 0.212 & 0.372 & 0.1914 & 0.0066 & 0.0151 & 0.1675 & 0.0141 & 0.0185 & 0.1816 & 0.0047 & 0.0130 & 0.1545 & 0.0134 & 0.0169 \\
  0.025 & 0.222 & 0.210 & 0.0769 & 0.0042 & 0.0168 & 0.1593 & 0.0181 & 0.0383 & 0.0786 & 0.0028 & 0.0114 & 0.1568 & 0.0176 & 0.0281 \\
  0.025 & 0.240 & 0.418 & 0.1885 & 0.0058 & 0.0111 & 0.1780 & 0.0133 & 0.0160 & 0.1801 & 0.0044 & 0.0106 & 0.1656 & 0.0128 & 0.0156 \\
  0.025 & 0.252 & 0.237 & 0.0737 & 0.0034 & 0.0120 & 0.1696 & 0.0175 & 0.0319 & 0.0786 & 0.0024 & 0.0089 & 0.1745 & 0.0181 & 0.0262 \\
  0.025 & 0.253 & 0.147 & 0.0375 & 0.0028 & 0.0144 & 0.1530 & 0.0198 & 0.0609 & 0.0410 & 0.0018 & 0.0093 & 0.1601 & 0.0193 & 0.0405 \\
  0.025 & 0.287 & 0.167 & 0.0365 & 0.0023 & 0.0103 & 0.1669 & 0.0197 & 0.0499 & 0.0413 & 0.0016 & 0.0075 & 0.1814 & 0.0204 & 0.0382 \\
  0.040 & 0.133 & 0.352 & 0.1524 & 0.0055 & 0.0134 & 0.1295 & 0.0125 & 0.0162 & 0.1546 & 0.0041 & 0.0109 & 0.1280 & 0.0126 & 0.0152 \\
  0.040 & 0.273 & 0.469 & 0.2027 & 0.0053 & 0.0088 & 0.2038 & 0.0133 & 0.0150 & 0.1905 & 0.0041 & 0.0085 & 0.1876 & 0.0126 & 0.0146 \\
  0.040 & 0.287 & 0.269 & 0.0798 & 0.0031 & 0.0093 & 0.2027 & 0.0190 & 0.0293 & 0.0812 & 0.0022 & 0.0073 & 0.2002 & 0.0190 & 0.0255 \\
  0.040 & 0.353 & 0.581 & 0.2048 & 0.0046 & 0.0057 & 0.2244 & 0.0107 & 0.0113 & 0.1926 & 0.0038 & 0.0055 & 0.2077 & 0.0103 & 0.0111 \\
  0.040 & 0.370 & 0.342 & 0.0753 & 0.0022 & 0.0047 & 0.2288 & 0.0173 & 0.0215 & 0.0723 & 0.0018 & 0.0042 & 0.2139 & 0.0166 & 0.0202 \\
  0.040 & 0.371 & 0.214 & 0.0386 & 0.0016 & 0.0055 & 0.2186 & 0.0209 & 0.0364 & 0.0393 & 0.0012 & 0.0042 & 0.2155 & 0.0206 & 0.0304 \\
  0.040 & 0.380 & 0.148 & 0.0232 & 0.0013 & 0.0061 & 0.2102 & 0.0230 & 0.0588 & 0.0255 & 0.0009 & 0.0041 & 0.2231 & 0.0235 & 0.0422 \\
  0.040 & 0.421 & 0.385 & 0.0737 & 0.0020 & 0.0036 & 0.2416 & 0.0162 & 0.0188 & 0.0708 & 0.0016 & 0.0031 & 0.2268 & 0.0157 & 0.0178 \\
  0.060 & 0.180 & 0.346 & 0.1036 & 0.0032 & 0.0070 & 0.1641 & 0.0148 & 0.0178 & 0.1045 & 0.0025 & 0.0062 & 0.1616 & 0.0149 & 0.0173 \\
  0.060 & 0.479 & 0.273 & 0.0383 & 0.0012 & 0.0030 & 0.2622 & 0.0203 & 0.0276 & 0.0384 & 0.0009 & 0.0025 & 0.2563 & 0.0202 & 0.0253 \\
  0.060 & 0.491 & 0.190 & 0.0233 & 0.0009 & 0.0033 & 0.2617 & 0.0234 & 0.0428 & 0.0247 & 0.0007 & 0.0025 & 0.2702 & 0.0240 & 0.0355 \\
  0.060 & 0.543 & 0.483 & 0.0743 & 0.0017 & 0.0022 & 0.2751 & 0.0138 & 0.0147 & 0.0717 & 0.0014 & 0.0021 & 0.2609 & 0.0135 & 0.0145 \\
  0.060 & 0.633 & 0.243 & 0.0209 & 0.0006 & 0.0016 & 0.2863 & 0.0205 & 0.0288 & 0.0222 & 0.0005 & 0.0013 & 0.2958 & 0.0212 & 0.0263 \\
  0.080 & 0.456 & 0.704 & 0.2404 & 0.0044 & 0.0048 & 0.2650 & 0.0086 & 0.0089 & 0.2245 & 0.0038 & 0.0048 & 0.2451 & 0.0081 & 0.0086 \\
  0.080 & 0.617 & 0.538 & 0.0762 & 0.0015 & 0.0018 & 0.2935 & 0.0124 & 0.0129 & 0.0725 & 0.0013 & 0.0018 & 0.2752 & 0.0119 & 0.0127 \\
  0.080 & 0.619 & 0.348 & 0.0366 & 0.0009 & 0.0016 & 0.2960 & 0.0179 & 0.0207 & 0.0349 & 0.0008 & 0.0014 & 0.2767 & 0.0170 & 0.0194 \\
  0.080 & 0.799 & 0.438 & 0.0337 & 0.0007 & 0.0008 & 0.3128 & 0.0135 & 0.0144 & 0.0324 & 0.0006 & 0.0009 & 0.2950 & 0.0131 & 0.0143 \\
  0.080 & 0.818 & 0.310 & 0.0198 & 0.0005 & 0.0008 & 0.3227 & 0.0171 & 0.0203 & 0.0195 & 0.0004 & 0.0007 & 0.3122 & 0.0167 & 0.0195 \\
  0.125 & 0.588 & 0.825 & 0.2925 & 0.0047 & 0.0050 & 0.2876 & 0.0061 & 0.0063 & 0.2666 & 0.0041 & 0.0048 & 0.2609 & 0.0055 & 0.0061 \\
  0.125 & 0.797 & 0.657 & 0.0801 & 0.0014 & 0.0015 & 0.3179 & 0.0089 & 0.0091 & 0.0732 & 0.0012 & 0.0015 & 0.2873 & 0.0082 & 0.0090 \\
  0.125 & 1.032 & 0.544 & 0.0329 & 0.0006 & 0.0007 & 0.3319 & 0.0097 & 0.0100 & 0.0296 & 0.0005 & 0.0007 & 0.2952 & 0.0087 & 0.0097 \\
  0.125 & 1.056 & 0.391 & 0.0185 & 0.0004 & 0.0005 & 0.3491 & 0.0131 & 0.0141 & 0.0174 & 0.0003 & 0.0005 & 0.3228 & 0.0122 & 0.0144 \\
  0.175 & 1.029 & 0.778 & 0.0876 & 0.0013 & 0.0013 & 0.3242 & 0.0060 & 0.0061 & 0.0773 & 0.0011 & 0.0011 & 0.2846 & 0.0052 & 0.0053 \\
  0.175 & 1.045 & 0.294 & 0.0085 & 0.0002 & 0.0003 & 0.3235 & 0.0142 & 0.0150 & 0.0078 & 0.0002 & 0.0002 & 0.2939 & 0.0129 & 0.0141 \\
  0.175 & 1.365 & 0.488 & 0.0165 & 0.0003 & 0.0003 & 0.3447 & 0.0109 & 0.0111 & 0.0148 & 0.0003 & 0.0003 & 0.3072 & 0.0099 & 0.0107 \\
  0.250 & 1.332 & 0.661 & 0.0305 & 0.0005 & 0.0005 & 0.3126 & 0.0074 & 0.0075 & 0.0262 & 0.0004 & 0.0005 & 0.2673 & 0.0065 & 0.0065 \\
  0.250 & 1.761 & 0.599 & 0.0144 & 0.0002 & 0.0002 & 0.3183 & 0.0082 & 0.0083 & 0.0125 & 0.0002 & 0.0002 & 0.2744 & 0.0071 & 0.0071 \\
  0.450 & 2.275 & 0.715 & 0.0099 & 0.0002 & 0.0002 & 0.2104 & 0.0046 & 0.0046 & 0.0077 & 0.0001 & 0.0001 & 0.1638 & 0.0036 & 0.0036 \\
\end{tabular}
\end{table}

\begin{table}
\caption{
  \label{ratio_dp}
  Ratio of deuteron to proton cross sections and structure functions. 
  Uncertainties are shown without ($\Delta_{noa2}$) and with ($\Delta_{full}$)
  the contribution from radiative corrections of order $\alpha^2$.
} 
\begin{tabular}{c c c c c c c c c c}
$x$  & $Q^2$ & $\varepsilon$ & $\sigma^d/\sigma^p$ & $\Delta_{noa2}$ &  $\Delta_{full}$ & $F_2^d/F_2^p$ & $\Delta_{noa2}$ &  $\Delta_{full}$  \\
\hline
 0.0080 & 0.0332 & 0.3521 & 0.8952 & 0.0540 & 0.0758 & 0.8798 & 0.0562 & 0.0774 \\
 0.0080 & 0.0509 & 0.2500 & 0.9426 & 0.0682 & 0.1339 & 0.9170 & 0.0728 & 0.1363 \\
 0.0080 & 0.0883 & 0.1605 & 0.8812 & 0.0719 & 0.1587 & 0.8448 & 0.0806 & 0.1628 \\
 0.0125 & 0.0359 & 0.3791 & 0.8513 & 0.0495 & 0.0633 & 0.8367 & 0.0516 & 0.0649 \\
 0.0125 & 0.1092 & 0.1977 & 1.0555 & 0.0644 & 0.1558 & 1.0129 & 0.0773 & 0.1616 \\
 0.0175 & 0.0593 & 0.3692 & 0.9613 & 0.0358 & 0.0418 & 0.9413 & 0.0410 & 0.0464 \\
 0.0175 & 0.0977 & 0.3247 & 1.0040 & 0.0373 & 0.0499 & 0.9766 & 0.0463 & 0.0569 \\
 0.0175 & 0.1553 & 0.2186 & 1.0137 & 0.0365 & 0.0598 & 0.9768 & 0.0519 & 0.0703 \\
 0.0250 & 0.0677 & 0.5390 & 0.9600 & 0.0298 & 0.0330 & 0.9452 & 0.0333 & 0.0361 \\
 0.0250 & 0.1085 & 0.3985 & 0.9840 & 0.0258 & 0.0282 & 0.9621 & 0.0338 & 0.0357 \\
 0.0250 & 0.2285 & 0.2567 & 0.9832 & 0.0247 & 0.0272 & 0.9529 & 0.0389 & 0.0405 \\
 0.0350 & 0.1022 & 0.5321 & 0.9646 & 0.0260 & 0.0274 & 0.9484 & 0.0306 & 0.0318 \\
 0.0350 & 0.1362 & 0.4030 & 0.9909 & 0.0254 & 0.0276 & 0.9689 & 0.0335 & 0.0352 \\
 0.0350 & 0.1765 & 0.2981 & 0.9942 & 0.0302 & 0.0408 & 0.9656 & 0.0416 & 0.0498 \\
 0.0350 & 0.3099 & 0.2974 & 0.9612 & 0.0221 & 0.0232 & 0.9375 & 0.0322 & 0.0329 \\
 0.0500 & 0.1713 & 0.5122 & 0.9732 & 0.0210 & 0.0210 & 0.9593 & 0.0251 & 0.0252 \\
 0.0500 & 0.3712 & 0.4090 & 0.9445 & 0.0199 & 0.0200 & 0.9273 & 0.0262 & 0.0262 \\
 0.0500 & 0.4644 & 0.2699 & 0.9897 & 0.0222 & 0.0232 & 0.9651 & 0.0330 & 0.0337 \\
 0.0700 & 0.2195 & 0.5146 & 0.9299 & 0.0213 & 0.0216 & 0.9140 & 0.0266 & 0.0268 \\
 0.0700 & 0.2692 & 0.4810 & 0.9576 & 0.0198 & 0.0199 & 0.9426 & 0.0247 & 0.0247 \\
 0.0700 & 0.4464 & 0.6947 & 0.9341 & 0.0193 & 0.0193 & 0.9253 & 0.0212 & 0.0212 \\
 0.0700 & 0.5624 & 0.4355 & 0.9641 & 0.0208 & 0.0210 & 0.9474 & 0.0267 & 0.0268 \\
 0.0700 & 0.6390 & 0.2965 & 0.9894 & 0.0225 & 0.0234 & 0.9670 & 0.0316 & 0.0323 \\
 0.0900 & 0.2936 & 0.4363 & 0.9530 & 0.0197 & 0.0197 & 0.9361 & 0.0259 & 0.0259 \\
 0.0900 & 0.4998 & 0.4794 & 0.9419 & 0.0187 & 0.0187 & 0.9310 & 0.0215 & 0.0215 \\
 0.0900 & 0.6328 & 0.5494 & 0.9384 & 0.0200 & 0.0200 & 0.9256 & 0.0238 & 0.0238 \\
 0.0900 & 0.7947 & 0.3526 & 0.9627 & 0.0212 & 0.0214 & 0.9436 & 0.0285 & 0.0286 \\
 0.1100 & 0.5649 & 0.8075 & 0.9197 & 0.0183 & 0.0183 & 0.9147 & 0.0190 & 0.0190 \\
 0.1100 & 0.8072 & 0.5470 & 0.9274 & 0.0187 & 0.0187 & 0.9162 & 0.0217 & 0.0217 \\
 0.1100 & 1.0138 & 0.4026 & 0.9491 & 0.0211 & 0.0211 & 0.9324 & 0.0269 & 0.0269 \\
 0.1400 & 0.4647 & 0.6166 & 0.9187 & 0.0182 & 0.0182 & 0.9099 & 0.0202 & 0.0202 \\
 0.1400 & 0.6204 & 0.8495 & 0.9080 & 0.0187 & 0.0187 & 0.9045 & 0.0190 & 0.0190 \\
 0.1400 & 0.8409 & 0.6831 & 0.9045 & 0.0191 & 0.0191 & 0.8967 & 0.0206 & 0.0206 \\
 0.1400 & 1.0623 & 0.5063 & 0.8971 & 0.0181 & 0.0181 & 0.8850 & 0.0218 & 0.0218 \\
 0.1800 & 0.5036 & 0.6698 & 0.9202 & 0.0183 & 0.0183 & 0.9136 & 0.0194 & 0.0194 \\
 0.1800 & 0.7065 & 0.9009 & 0.8918 & 0.0189 & 0.0189 & 0.8901 & 0.0190 & 0.0190 \\
 0.1800 & 0.9981 & 0.5371 & 0.8861 & 0.0176 & 0.0176 & 0.8802 & 0.0185 & 0.0185 \\
 0.1800 & 1.3393 & 0.5233 & 0.8947 & 0.0179 & 0.0179 & 0.8853 & 0.0202 & 0.0202 \\
 0.2250 & 0.7407 & 0.9170 & 0.8775 & 0.0173 & 0.0173 & 0.8765 & 0.0173 & 0.0173 \\
 0.2250 & 1.0817 & 0.5354 & 0.8625 & 0.0171 & 0.0171 & 0.8583 & 0.0176 & 0.0176 \\
 0.2250 & 1.3655 & 0.6727 & 0.8743 & 0.0177 & 0.0177 & 0.8700 & 0.0182 & 0.0182 \\
 0.2250 & 1.6733 & 0.5761 & 0.8817 & 0.0180 & 0.0180 & 0.8775 & 0.0185 & 0.0185 \\
 0.2750 & 0.7749 & 0.9313 & 0.8371 & 0.0176 & 0.0176 & 0.8367 & 0.0176 & 0.0176 \\
 0.2750 & 1.2362 & 0.8579 & 0.8305 & 0.0171 & 0.0171 & 0.8303 & 0.0171 & 0.0171 \\
 0.2750 & 1.7101 & 0.5221 & 0.8667 & 0.0169 & 0.0169 & 0.8653 & 0.0170 & 0.0170 \\
 0.3500 & 1.3006 & 0.5544 & 0.8340 & 0.0153 & 0.0153 & 0.8340 & 0.0153 & 0.0153 \\
 0.3500 & 1.7109 & 0.5575 & 0.8279 & 0.0153 & 0.0153 & 0.8279 & 0.0153 & 0.0153 \\
 0.3500 & 2.1840 & 0.6967 & 0.8243 & 0.0167 & 0.0167 & 0.8243 & 0.0167 & 0.0167 \\
 0.4500 & 2.3660 & 0.7325 & 0.7881 & 0.0144 & 0.0144 & 0.7881 & 0.0144 & 0.0144 \\
\end{tabular}
\end{table}

\end{widetext}


\begin{thebibliography}{99}
\bibitem{vtmail}
The numerical values of all data are available at hallcweb.jlab.org/DISdata.
\end{thebibliography}
\end{document}